\documentstyle[psfig,epsf,rotate,aps]{revtex}
\begin{document}

\title{The principal eigenvector of contact matrices
and hydrophobicity profiles in proteins}

\author{
Ugo~Bastolla$^{1}$,
Markus~Porto$^{2,3}$\footnote{Present address:
Institut~f\"ur~Festk\"orperphysik, Technische~Universit\"at~Darmstadt,
Hochschulstr.~8,  64289~Darmstadt, Germany},
H.~Eduardo~Roman$^{4}$ and Michele~Vendruscolo$^{5}$}

\address{
$^{1}$Centro~de~Astrobiolog{\'\i}a~(INTA-CSIC),
             28850~Torrej\'on~de~Ardoz, Spain \\
$^{2}$Max-Planck-Institut~f\"ur~Physik~komplexer~Systeme,
             N\"othnitzer~Stra{\ss}e~38, 01187~Dresden, Germany \\
$^{3}$Institut~f\"ur~Theoretische~Physik,
             Technische~Universit\"at~Dresden, 01062~Dresden, Germany \\
$^{4}$Dipartimento~di~Fisica and INFN, Universit\`a~di~Milano,
             Via~Celoria~16, 20133~Milano, Italy \\
$^{5}$Department~of~Chemistry, University~of~Cambridge,
             Lensfield~Road, Cambridge CB2~1EW, UK
}

\date{\today}
\maketitle

\begin{abstract}
With the aim to study the relationship between protein sequences and their
native structures, we adopt vectorial representations for both sequence and
structure. The structural representation is based on the Principal Eigenvector
of the fold's contact matrix (PE). As recently shown, the latter encodes
sufficient information for reconstructing the whole contact matrix. The
sequence is represented through a Hydrophobicity Profile (HP), using a
generalized hydrophobicity scale that we obtain from the principal eigenvector
of a residue-residue interaction matrix and denote it as {\em interactivity}
scale. Using this novel scale, we define the optimal HP of a protein fold, and
predict, by means of stability arguments, that it is strongly correlated with
the PE of the fold's contact matrix. This prediction is confirmed through an
evolutionary analysis, which shows that the PE correlates with the HP of each
individual sequence adopting the same fold and, even more strongly, with the
average HP of this set of sequences. Thus, protein sequences evolve in such a
way that their average HP is close to the optimal one, implying that neutral
evolution can be viewed as a kind of motion in sequence space around the
optimal HP. Our results indicate that the correlation coefficient between
$N$-dimensional vectors constitutes a natural metric in the vectorial space in
which we represent both protein sequences and protein structures, which we call
Vectorial Protein Space. In this way, we define a unified framework for
sequence to sequence, sequence to structure, and structure to structure
alignments. We show that the interactivity scale is nearly optimal both for the
comparison of sequences with sequences and sequences with structures.
\end{abstract}

\vspace{0.5cm}
\centerline{\textbf{Keywords:}
Vectorial representation of proteins, Protein folding, Hydrophobicity,}
\centerline{Contact maps, Vectorial Protein Space.}

\clearpage

\textwidth 16cm
\twocolumn

\section*{INTRODUCTION}

The information contained in protein sequences can be represented by intrinsic
profiles, such as hydrophobicity \cite{Kyte82,Sweet83}, charge, and secondary
structure propensities. Structural information can also be reduced to
one-dimensional profiles describing structural and chemical properties of the
amino acids \cite{Bowie91}, including secondary structure and solvent
accessibility \cite{Rost95}. It has been shown that the Hydrophobicity Profile
(HP) is correlated with the solvent accessibility of the native structure
\cite{Bowie90}, indicating that sequence and structure profiles are intimately
related \cite{Willmans93,Huang95}.

In order to gain further insight into the sequence-structure relationship, we
represent both protein sequences and protein structures as vectors in an
$N$-dimensional space, denoted as Vectorial Protein Space, and study their
mutual relationship. In this work, we adopt a structural representation of
proteins based on the Principal Eigenvector of the native contact matrix (PE).
The PE has already been used as an indicator of protein topology, in particular
as a mean of identifying structural domains \cite{Holm94} and clusters of amino
acids with special structural significance \cite{Kannan99,Kannan00}. We have
recently shown that knowledge of the PE suffices to reconstruct the complete
contact matrix of single-domain globular proteins \cite{Porto04}.

Protein sequences are here represented through their HP \cite{Sweet83}. We
introduce a new generalized hydrophobicity scale that is based on the principal
eigenvector of a residue-residue interaction matrix. This scale, which we call
{\em interactivity}, correlates strongly with empirical hydrophobicity scales,
even if it incorporates also other interactions besides the hydrophobic effect.
The inclusion of the native fold's effective free energy in the new
interactivity scale is crucial for deducing the relationship between PE and HP.

We define the optimal HP for a given fold and show that a strong correlation
can be expected between this optimal HP and the PE. Thus, the optimal HP
carries the ``hydrophobic fingerprint'' of a protein fold. While we do not
expect that the optimal HP is realized in protein evolution, we do expect that
protein evolution is confined in a region of sequence space centered around the
optimal HP, so that the evolutionary average of the HP over sequences sharing
the same fold should more strongly resemble the optimal HP.

To test this expectation, we performed an evolutionary analysis using two kinds
of protein sets: (i)~the PFAM \cite{Bateman00} and FSSP \cite{Holm96}
databases, containing sequences of protein families adopting the same fold, and
(ii)~the SCN database, containing sequences of seven protein families obtained
through the Structurally Constrained Neutral (SCN) model of evolution
\cite{Bastolla03a,Bastolla02,Bastolla03b} that imposes the conservation of the
thermodynamic stability of the native fold. For all three databases we found
that the HP of individual sequences are positively correlated to the PE, yet
the evolutionary average of the HP is more strongly correlated to it than each
sequence individually. This finding suggests that the HP of a protein sequence
encodes a large part of the information on the PE of the native contact matrix
and hence on the native structure. The correlation between the average HP and
the PE allows deriving a site-specific pattern of sequence conservation, which
can be used in evolutionary studies.

Representing protein structures in the same vector space as protein sequences
permits to define a distance measure, based on the correlation coefficient,
which is applicable to sequence-sequence, structure-structure, as well as
sequence-structure comparisons. This definition provides a unified framework
for all kinds of protein alignments. In this context, we compared the
similarity score based on the Hydrophobicity Correlation (HC), already proposed
by Sweet and Eisenberg \cite{Sweet83}, with the widely used BloSum
\cite{Henikoff92} score, using as benchmark a set of distantly related proteins
sharing the TIM barrel fold \cite{Nagano02}. We also derived independently
optimal parameters for sequence-structure and for sequence-sequence
comparisons. In both cases, the interactivity parameters are almost identical
with the optimal ones.

%%%%%%%%%%%%%%%%%%%%%%%%%%%%%%%%%%%%%%%%%%%%%%%%%%%%%%%%%%%%%%%%%%%%%%%%%%%%

\section*{METHODS}

\subsection*{The contact matrix and its spectral properties}

The contact matrix ${\mathbf{C}}$ is a binary matrix, with elements $C_{ij} =
1$ if amino acids at positions $i$ and $j$ are in contact, and $0$ otherwise.
Only residues separated by at least three positions along the sequence are
considered in contact, so that $C_{ij} = 0$ for all $i$, $j$ with $|i-j| < 3$.
Two residues are considered to be in contact if any two of their heavy atoms
are closer than $4.5 \, \mbox{\AA}$ in space. Therefore, the contact condition
depends on the size of the amino acids at positions $i$ and $j$.

Since the contact matrix is an $N\times N$ symmetric matrix, it has $N$ real
eigenvalues $\lambda_{\alpha}$, $\alpha = 1, \ldots, N$, which we rank in
decreasing order. The corresponding eigenvectors $\{ {\mathbf{c}}^{(\alpha)}
\}$ form an orthonormal system.\footnote{In the following, we indicate vectors
and matrices using \textbf{bold face} letters.}

\subsection*{Effective connectivity and principal eigenvector}

 From the contact matrix ${\mathbf{C}}$, we define a vectorial representation
${\mathbf{c}}({\mathbf{C}})$, which we call the effective connectivity, such
that positions $i$ with large $c_i({\mathbf{C}})$ are in contact with as many
as possible positions $j$ with large $c_j({\mathbf{C}})$. This self-consistent
definition can be formally expressed through the condition that the vector
${\mathbf{c}}({\mathbf{C}})$ maximizes the quadratic form
\begin{equation}\label{eq:quad}
Q = \sum_{ij} C_{ij} \, c_i \, c_j \, ,
\end{equation}
with the normalization condition $\sum_{i=1}^N c_i^2 = 1$. The solution of this
maximization problem is the Principal Eigenvector (PE) of the contact matrix
corresponding to the largest eigenvalue $\lambda_1$,
${\mathbf{c}}({\mathbf{C}}) = {\mathbf{c}}^{(1)}$. In the following, we denote
the PE by ${\mathbf{c}}$ instead of ${\mathbf{c}}^{(1)}$.

 From this maximization property, it follows that the largest positive
eigenvalue $\lambda_1$ has a value ranging between the average number of
contacts per residue, $\sum_{ij} C_{ij}/N$, and the maximal number of contacts
of any given residue, $\max_i \left( \sum_j C_{ij} \right)$ \cite{Bollobas98}.
In addition, since all elements of ${\mathbf{C}}$ are positive or zero, the PE
has all components of the same sign or zero. We choose by convention the
positive sign.

PE components are zero only for residues that do not form contacts with
residues with non-vanishing PE. Vanishing PE components may indicate a domain
decomposition of the protein structure. The components of the PE outside the
main domain are never exactly zero, but their value is much smaller than inside
the main domain. The algorithm for automatic domain decomposition proposed by
Holm and Sander \cite{Holm94} is based on a similar idea. For this reason,
multi-domain proteins are expected to have a larger variance of their PE
components than single-domain ones. A study of the Protein Data Bank (PDB) has
confirmed this expectation (data not shown). Therefore, we use as a signature
for single-domain proteins a small relative variance of the PE, the latter
defined as
\begin{equation}\label{eq:B}
B \equiv
\frac{\sum_i \left( c_i - \big< c \big> \right)^2}%
{N \big< c \big>^2} =
\frac{1-N \big< c \big>^2}{N \big< c \big>^2} \, ,
\end{equation}
where $\big<c \big> = N^{-1} \sum_i c_i$.

\subsection*{Similarity measure}

Both the PE and the HP are vectors in an $N$-dimensional space, where $N$ is
the number of amino acids. The most natural way of defining a similarity
measure in this space is through the normalized scalar product: The cosine of
the angle $\alpha$ between two vectors ${\mathbf{x}}$ and ${\mathbf{y}}$ is
defined as $\cos \alpha = \sum x_i y_i/\sqrt{\sum_i x_i^2 \sum_i y_i^2}$. This
quantity, however, is strongly dependent on the average value of the two
vectors, $\big< x \big> = N^{-1} \sum_i x_i$ and $\big< y \big> = N^{-1} \sum_i
y_i$. It is therefore more convenient to use as similarity measure Pearson's
correlation coefficient $r({\mathbf{x}},{\mathbf{y}})$, which is the normalized
scalar product of the vectors with components $x_i - \big< x \big>$ and $y_i -
\big< y \big>$ and is defined as
\begin{equation}\label{eq:corr}
r({\mathbf{x}},{\mathbf{y}}) \equiv
\frac{\sum_i \left( x_i - \big< x \big> \right)
\left( y_i - \big< y \big> \right)}
{\sqrt{\left[ \sum_i \left( x_i - \big< x \big> \right)^2 \right]
\left[ \sum_i \left( y_i - \big< y \big> \right)^2 \right]}} \, .
\end{equation}

\subsection*{Effective folding free energy}

We will use in the following a simple model of protein thermodynamics. In this
model, the free energy of a sequence ${\mathbf{A}}$ folded into a contact map
${\mathbf{C}}$ is approximated by an effective contact free energy function
$E({\mathbf{A}},{\mathbf{C}})$,
\begin{equation}\label{eq:energy}
\frac{E({\mathbf{A}},{\mathbf{C}})}{k_{\mathrm{B}} T} =
\sum_{i < j} C_{ij} \, U(A_i, A_j) \, ,
\end{equation}
where ${\mathbf{U}}$ is a $20\times 20$ symmetric matrix with $U(a,b)$
representing the effective interaction, in units of $k_{\mathrm{B}} T$, of
amino acids $a$ and $b$ when they are in contact. We use the interaction matrix
derived by Bastolla \textit{et al.} \cite{Bastolla01}.

\subsection*{The SCN model of protein evolution}

The Structurally Constrained Neutral (SCN) model
\cite{Bastolla03a,Bastolla02,Bastolla03b} is based on mutations and purifying
selection. Selection is imposed requiring that the folding free energy and the
normalized energy gap of the native structure, calculated through the effective
free energy function Eq.~(\ref{eq:energy}) \cite{Bastolla01}, are above
predetermined thresholds. The model reproduces the main qualitative features of
protein sequence evolution and allows structurally based sequence conservation
for specific protein folds to be predicted
\cite{Bastolla03a,Bastolla02,Bastolla03b}. The conservation values estimated
through the SCN model are in agreement with those measured in the corresponding
FSSP family, with the exception of functionally constrained positions, which
are not conserved in our model \cite{Bastolla03a}.

\subsection*{Sequence Databases}

We studied in detail seven protein folds with different length $N$: The TIM
barrel ($N=247$, PDB code \texttt{7tim\_A}), the ubiquitin conjugating enzyme
($N=160$, PDB code \texttt{1u9a\_A}), myoglobin ($N=151$, PDB code
\texttt{1a6g}), lysozyme ($N=129$, PDB code \texttt{3lzt}), ribonuclease
($N=124$, PDB code \texttt{7rsa}), cytochrome~c ($N=82$, PDB code
\texttt{451c}), rubredoxin ($N=53$, mesophilic: PDB code \texttt{1iro};
thermophilic: PDB code \texttt{1brf\_A}).

For each fold, we collected three families of aligned sequences expected to
belong to the same structural class: (i)~The PFAM family \cite{Bateman00},
grouped through sequence comparison techniques. (ii)~The FSSP family
\cite{Holm96}, grouped through structure comparison techniques. (iii)~The SCN
family \cite{Bastolla03a,Bastolla02,Bastolla03b}, obtained by simulating
molecular evolution with the constraint that the thermodynamic stability of the
native structure must be conserved.

\subsection*{Optimization of hydrophobicity parameters}

In this work, we use generalized hydrophobicity parameters that are obtained
from the principal eigenvector of the interaction matrix represented in
Eq.~(\ref{eq:energy}) \cite{Bastolla01,Bastolla00}. We call this set of
parameters the {\em interactivity} parameters, and we use them to obtain a
vectorial representation of protein sequences. Protein structures are also
represented as vectors through the Principal Eigenvector (PE) of their contact
matrix. We will show here that the interactivity parameters simultaneously
confer high similarity score, \textit{i.e.} high correlation coefficient,
(a)~to pairs of vectors representing a protein sequence and its native
structure, and (b)~to pairs of distantly related sequences sharing the same
structure (see below).

We also determined two new sets of 20 normalized parameters by maximizing the
two scores (a) and (b) respectively, averaged over a training set of proteins.
For the optimization we adopted an \textit{ad hoc} method based on the fact
that a generic parameter set can be written as a linear combination of the $20$
eigenvectors of the interaction matrix ${\mathbf{U}}$. For parameter sets
consisting of a single eigenvector, the score is large for the principal
eigenvector, medium for a few eigenvectors and low for all other ones.
Therefore, it is reasonable to expect that the optimal set will be a
combination of a small number of eigenvectors. Our method works in three steps.
(i)~For all $19\cdot 18\cdot 17/6=969$ combinations of four eigenvectors of
${\mathbf{U}}$ (always including the principal one), we maximized the average
score as a function of the three corresponding coefficients (one is fixed by
the normalization condition), using exact enumeration with large steps. (ii)~We
selected the six combinations of four eigenvectors giving the largest scores
and we optimized the coefficients using smaller steps. (iii)~We checked that
addition of each of the remaining eigenvectors did not improve the result
significantly.

In case (a), we used as training set the single-domain globular proteins
described below. In case (b), the similarity score
$r({\mathbf{h}},{\mathbf{h}}')$ was calculated for pairs of distantly related
sequences with the TIM barrel fold. In both cases, the optimization was also
performed with standard Monte Carlo techniques, yielding a very similar result.
In case (b), we did not optimize the placement of gaps at the same time, but we
used gapped alignments obtained through structural superposition downloaded
from the Dali server (\texttt{http://www.ebi.ac.uk/dali/fssp/}).

\subsection*{PDB set}

We computed the correlation between HP and PE for a subset of non-redundant
single-domain globular proteins ($404$ single-domain structures of length
$N\leq 200$). We tested the condition of globularity by imposing that the
fraction of contacts per residue was larger than a length dependent threshold,
$N_{\mathrm{c}}/N > 3.5 + 7.8 N^{-1/3}$. This functional form represents the
scaling of the number of contacts in globular proteins as a function of chain
length (the factor $N^{-1/3}$ comes from the surface to volume ratio), and the
two parameters were chosen so to eliminate outliers with respect to the general
trend, which are mainly non-globular structures. The condition of being
single-domain was imposed by checking that the variance of the PE components,
Eq.~(\ref{eq:B}), was $B < 1.5$.

%%%%%%%%%%%%%%%%%%%%%%%%%%%%%%%%%%%%%%%%%%%%%%%%%%%%%%%%%%%%%%%%%%%%%%%%%%%%

\section*{RESULTS}

\subsection*{Optimal HP}

We investigate the relationship between protein sequence and structure using an
effective folding free energy function, based on the pairwise interaction
matrix ${\mathbf{U}}$, see Eq.~(\ref{eq:energy}). This interaction matrix can
be written in spectral form as $U(a,b) = \sum_{\alpha=1}^{20} \epsilon_{\alpha}
\, u^{(\alpha)}(a) \, u^{(\alpha)}(b)$, where $\epsilon_{\alpha}$ are the
eigenvalues, ranked by their absolute value, and ${\mathbf{u}}^{(\alpha)}$ are
the corresponding eigenvectors. The contribution of the principal eigenvector
${\mathbf{u}}^{(1)}$ to the spectral decomposition, $\epsilon_1 \, u^{(1)}(a)
\, u^{(1)}(b)$, with $\epsilon_1<0$, constitutes the main component of the
interaction matrix. It has correlation coefficient $0.81$ with the elements
$U(a,b)$ of the full matrix. Thus, an approximated effective energy function,
yielding a good approximation to the full contact energy Eq.~(\ref{eq:energy}),
can be obtained as
\begin{equation}\label{eq:hydro}
\frac{H({\mathbf{A}},{\mathbf{C}})}{k_{\mathrm{B}} T} \equiv
\epsilon_1 \sum_{i < j} C_{ij} \, h(A_i) \, h(A_j)\, ,
\end{equation}
where the set of parameters $h(a)=u^{(1)}(a)$ coincide with the main
eigenvector of the interaction matrix.

It has been shown by Li \textit{et al.} \cite{Li97} that the eigenvector of the
Miyazawa and Jernigan contact interaction matrix \cite{Miyazawa85}
corresponding to the largest (in absolute value) eigenvalue is related to
hydrophobicity, having a correlation coefficient of $0.77$ with an empirical
hydrophobicity scale. For the interaction matrix used in this study
\cite{Bastolla01}, the corresponding eigenvector, ${\mathbf{u}}^{(1)}$, has a
correlation coefficient of $0.85$ with the Fauchere and Pliska hydrophobicity
scale \cite{Fauchere83} (cf.\ Table~\ref{tab:hydro}).

We call the main eigenvector of the interaction matrix, $h(a)\equiv
u^{(1)}(a)$, the {\em interactivity} of the corresponding residues $a$. These
parameters are based also on other interactions besides the hydrophobic effect,
{\it e.g.} electrostatic interactions, but the hydrophobicity is their main
component, as indicated by the strong correlation between the interactivity and
the Fauchere and Pliska hydrophobicity scale. Therefore, we will call the
$N$-dimensional vector ${\mathbf{h}}({\mathbf{A}})$ the Hydrophobicity Profile
of sequence ${\mathbf{A}}$, abbreviated in the following as HP.

%%%%%%%%%%%%%%%%%%%%%%% Tab. 1 %%%%%%%%%%%%%%%%%%%%%%%%%%%%%%%

\begin{table}
\begin{center}
\begin{tabular}{|c|r|r|r|r|}\hline
A  & $  0.1366$ & $ 0.0728$ & $ 0.1510$ & $ 0.31$ \\
E  & $ -0.0484$ & $-0.0295$ & $-0.0639$ & $-0.64$ \\
Q  & $  0.0325$ & $ 0.0126$ & $ 0.0246$ & $-0.22$ \\
D  & $ -0.1233$ & $-0.0552$ & $ 0.0047$ & $-0.77$ \\
N  & $ -0.0345$ & $-0.0390$ & $ 0.0381$ & $-0.60$ \\
L  & $  0.4251$ & $ 0.3819$ & $ 0.3926$ & $ 1.70$ \\
G  & $ -0.0464$ & $-0.0589$ & $ 0.0248$ & $ 0.00$ \\
K  & $ -0.0101$ & $-0.0053$ & $-0.0158$ & $-0.99$ \\
S  & $ -0.0433$ & $-0.0282$ & $ 0.0040$ & $-0.04$ \\
V  & $  0.4084$ & $ 0.2947$ & $ 0.3997$ & $ 1.22$ \\
R  & $  0.0363$ & $ 0.0394$ & $-0.0103$ & $-1.01$ \\
T  & $  0.0589$ & $ 0.0239$ & $ 0.1462$ & $ 0.26$ \\
P  & $  0.0019$ & $-0.0492$ & $ 0.0844$ & $ 0.75$ \\
I  & $  0.4172$ & $ 0.3805$ & $ 0.4238$ & $ 1.80$ \\
M  & $  0.1747$ & $ 0.1613$ & $ 0.2160$ & $ 1.23$ \\
F  & $  0.4076$ & $ 0.4201$ & $ 0.3455$ & $ 1.79$ \\
Y  & $  0.3167$ & $ 0.3113$ & $ 0.2998$ & $ 0.96$ \\
C  & $  0.2745$ & $ 0.3557$ & $ 0.3222$ & $ 1.54$ \\
W  & $  0.2362$ & $ 0.4114$ & $ 0.2657$ & $ 2.25$ \\
H  & $  0.0549$ & $ 0.0874$ & $ 0.1335$ & $ 0.13$ \\
\hline
\end{tabular}
\caption{Interactivity scales derived in this work. In column~2, the parameters
are obtained from the components of the principal eigenvector of the contact
interaction matrix \protect\cite{Bastolla01} (see 'Optimal HP'). In column~3,
the parameters are obtained maximizing the mean $r({\mathbf{c}},{\mathbf{h}})$
over single-domain globular proteins (see 'Parameter optimization'). In
column~4, the parameters are obtained by maximizing the mean
$r({\mathbf{h}},{\mathbf{h}}')$ over pairs of sequences sharing the TIM barrel
fold (see 'Sequence comparison'). Column~5 shows the hydrophobicity scale
determined by Fauchere and Pliska \protect\cite{Fauchere83}.}
\label{tab:hydro}
\end{center}
\end{table}

By comparing Eq.~(\ref{eq:hydro}) to the definition of the PE ${\mathbf{c}}$ of
the contact matrix ${\mathbf{C}}$, Eq.~(\ref{eq:quad}), one sees that the HP
satisfying $h(A_i) = {\mathrm{const}} \times c_i$ minimizes the energy
$H({\mathbf{A}},{\mathbf{C}})$ for a fixed value of the sum $\sum_{i=1}^N
\left[ h(A_i) \right]^2$. We define the optimal HP of the contact matrix
${\mathbf{C}}$ as the vector ${\mathbf{h}}$ minimizing
$H({\mathbf{A}},{\mathbf{C}})$ with the constraints of fixed $\sum_i \left[
h(A_i) \right]^2$ and fixed average hydrophobicity $\big< h \big> = N^{-1}
\sum_i h(A_i)$. The latter condition is imposed because, if the average
hydrophobicity is large, the hydrophobic energy will be low not only for the
native contact map, but also for alternative, misfolded structures. Thus $\big<
h \big>$ has to be reduced in order to obtain a sequence with a large
normalized energy gap and a well correlated free energy landscape, which is a
requisite for fast folding and thermodynamic stability, as well as evolutionary
stability \cite{BW87,Goldstein92,Abkevich94,Gutin95,Klimov96,Bastolla00}.

The exact solution of this optimization problem as a function of $\big< h
\big>$ is involved. Here we report only its qualitative features: The
normalized scalar product of the optimal HP and the PE equals one if the
average hydrophobicity satisfies $\big< h \big> = \big< h \big>_0 \equiv \big<
c \big> \sqrt{\sum_i h_i^2}$, which is the value where the energy
$H({\mathbf{A}},{\mathbf{C}})$ is minimal, and then decreases proportionally to
$\left( \big< h \big>-\big< h \big>_0 \right)^2$.

On the basis of the interactivity scale, protein sequences in the PDB have an
average hydrophobicity $\big< h \big> \simeq 0.77 \big< h \big>_0$, a value at
which the optimal HP is still almost parallel to the PE: The scalar product
between PE and optimal HP was larger than $0.85$ in all cases in which we
calculated it. It is interesting that $\big< h \big>$ in globular proteins in
the PDB is always smaller than $\big< h \big>_0$. A larger value of $\big< h
\big>$ would yield a lower hydrophobic free energy
$H({\mathbf{A}},{\mathbf{C}})$, but it would decrease the energy gap, with the
consequence of making folding less efficient.

Since all its components are positive, the PE has a large average value $\big<
c \big>$, which contributes significantly to the scalar product. Therefore, it
is more convenient to measure the similarity between PE and HP through their
correlation coefficient $r({\mathbf{c}},{\mathbf{h}})$ (see Methods).
The correlation coefficient between the optimal HP and the PE is
expected to be close to one. This relationship between the PE and the optimal
HP is very useful to investigate the relationship between sequences and
structures, and it provides a new perspective on protein evolution.

\subsection*{Evolutionary average of hydrophobicity profiles}

We do not expect sequences resulting from an evolutionary process to have
optimal HP. In a population with $M$ individuals, natural selection is only
able to fix advantageous mutations whose selective advantage is at least of
order $1/M$. Variants with lower (positive) selective advantage are
effectively neutral and evolve by random genetic drift \cite{Kimura83}. When
the thermodynamic stability of the protein is large, mutations improving
stability are less likely to occur. Therefore one expects that protein
stability does not overcome a typical value that depends on the population
size, the selection strength, and the mutation rate. Moreover, selection for
proper function places constraints on key amino acids and may prevent
proteins to reach thermodynamically optimal sequences.

Nevertheless, we do expect a positive correlation between HP and PE, since
this gives an important contribution to the stability of the native structure.
Therefore, we predict that protein evolution visits a region in the
hydrophobicity space centered about the optimal HP. To test this hypothesis,
we average the HP over sets of aligned sequences derived from the PFAM, the
FSSP, and the SCN databases (see Methods).

\begin{figure}[ht]
\center{\leavevmode
\epsfxsize 7.6cm
\epsffile{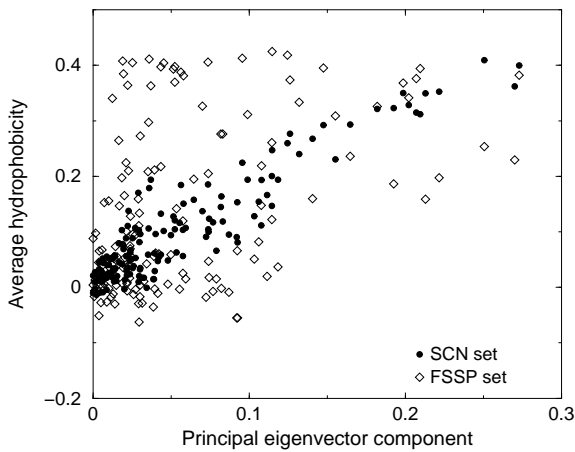}
}
\caption{Scatter plot of PE components, $c_i$, and HP components, $h(A_i)$, the
latter ones averaged over two sets of sequences with the myoglobin fold: The
SCN set obtained through simulated evolution (circles) and the FSSP set
obtained through structure comparison (diamonds). Correlation coefficients are
$0.96$ and $0.46$, respectively.}
\label{fig:myog_w}
\end{figure}

Table~\ref{tab:summary} presents the correlation coefficients
$r({\mathbf{c}},{\mathbf{h}})$ between the PE of the native structure and five
HP, for seven protein folds of various lengths. First, we calculate
$r({\mathbf{c}},{\mathbf{h}})$ using the HP of the native sequence in the PDB.
The mean value is $0.47$ (fold average). The probability that this correlation
arises by chance is comprised between less than $10^{-4}$ in the case of
rubredoxin ($N=53$) and less than $10^{-13}$ in the case of the TIM barrel
($N=247$). Second, we average $r({\mathbf{c}},{\mathbf{h}})$ over all
homologous sequences in the same PFAM class, obtaining a fold average of
$0.45$, very similar to the previous one. The same procedure similarly gives a
mean correlation coefficient of $0.45$ with sequences in the same FSSP class,
and $0.45$ with sequences in the same SCN class (not shown). We then average
the HP over sequences in the same family and use it to calculate
$r({\mathbf{c}},{\mathbf{h}})$, finding significantly larger values. The mean
$r({\mathbf{c}},{\mathbf{h}})$ is $0.57$ when the HP is averaged over a PFAM
class, $0.58$ when it is averaged over a FSSP class, and $0.96$ when it is
averaged over the set of sequences obtained through the SCN model. In
Fig.~\ref{fig:myog_w} we show a scatter plot of the PE versus the average HP
over the SCN set and the FSSP set for the myoglobin fold.

This analysis establishes an important result: The evolutionary-averaged HP
correlates with the PE better than the HP of each individual sequence. This
suggests that, although individual sequences are not optimal, the evolutionary
process moves around the optimal sequence in sequence space. This
interpretation is strongly supported by the results of the SCN model, for
which the average HP is very close to the optimal HP, although individual
sequences have $r({\mathbf{c}},{\mathbf{h}})$ not much different from those
of the PFAM and FSSP databases.

In order to get a deeper insight into the mechanism by which the SCN model
operates, we have simulated the evolution of the seven protein folds of
Table~\ref{tab:summary} by imposing the selective constraint either on (a)~the
normalized energy gap, or (b)~the folding free energy. The SCN model uses a
combination of both constraints. When selection is imposed only on the energy
gap, case (a), the folding free energy is higher and the mean correlation
coefficient decreases by around 15\%. This agrees with our argument that the
strong correlation between average HP and PE arises from minimization of the
native energy. In contrast, when only the folding free energy is tested to
accept mutations, case (b), the mean correlation coefficient increases only
slightly, but the energy does not change significantly with respect to the SCN
model. This is in line with our expectation that relaxing the constraint on the
energy gap increases the correlation between the PE and the optimal HP.

Last, we test the influence of the size of sequence database on the value of
the correlation coefficient. Whereas FSSP and PFAM families consist of a few
tens or hundreds of very correlated sequences, the number of sequences in an
SCN family is of tens of thousands. The correlation between the PE and the
average HP decreases if a smaller number of sequences is used to calculate the
average, and it becomes similar to the value obtained for the FSSP and PFAM
databases when we use only hundreds of sequences.

\subsection*{Parameters optimization}

In the SCN model we assume that Eq.~(\ref{eq:energy}) gives the exact free
energy of the system. From this assumption, we have obtained here the
interactivity parameters by diagonalizing the contact interaction matrix
${\mathbf{U}}$ \cite{Bastolla01}. However, Eq.~(\ref{eq:energy}) is only
approximate for real proteins. It is therefore possible that the correlation
between average HP and PE can be improved by using another set of 20
parameters.

To test this possibility, we first calculated the correlation coefficient of
the PE with the HP obtained using the hydrophobicity parameters measured by
Fauchere and Pliska \cite{Fauchere83}. The correlation coefficients between HP
and PE remain in this case very similar to those calculated above.

We then optimized the set of $20$ parameters by maximizing the mean correlation
coefficient $r({\mathbf{c}},{\mathbf{h}})$ over a subset of the PDB containing
single-domain globular structures (see Methods). The optimal parameters,
reported in Table~\ref{tab:hydro}, were very similar to the parameters obtained
from the principal eigenvector of the matrix ${\mathbf{U}}$ (correlation
coefficient $0.95$) and the mean $r({\mathbf{c}},{\mathbf{h}})$ so obtained was
less than 4\% better than the value previously obtained. Therefore, the
interactivity parameters that we are using in this study are almost optimal
from the point of view of maximizing $r({\mathbf{c}},{\mathbf{h}})$ for PDB
proteins.

\subsection*{Sequence comparison}

We have seen that the HP of a protein sequence is correlated with the PE of its
native structure. This implies that the HP of sequences folding into a similar
structure correlate with each other. Therefore, one can measure the similarity
of two protein sequences through their HP correlation (HC) score, defined as
\begin{equation}
{\mathrm{HC}}({\mathbf{A}},{\mathbf{A}}') \equiv
r({\mathbf{h}}({\mathbf{A}}),{\mathbf{h}}({\mathbf{A}}')) \, ,
\end{equation}
where the correlation $r$ is defined in Eq.~(\ref{eq:corr}) and
${\mathbf{h}}({\mathbf{A}})$ indicates the HP of sequence ${\mathbf{A}}$. This
sequence similarity measure was introduced by Sweet and Eisenberg
\cite{Sweet83}.

We compared this similarity score with the score obtained through the BloSum62
score matrix \cite{Henikoff92}, a substitution matrix commonly used in
bioinformatics applications. We applied the two scores to the family of
sequences sharing the TIM barrel fold, aligned through the structural alignment
algorithm Dali \cite{Holm96}. This family possesses many sequence pairs whose
relationship is very difficult to detect on the basis of the sequence alone,
even with the best current algorithms \cite{Nagano02}.

\begin{figure}[ht]
\center{\leavevmode
\epsfxsize 7.6cm
\epsffile{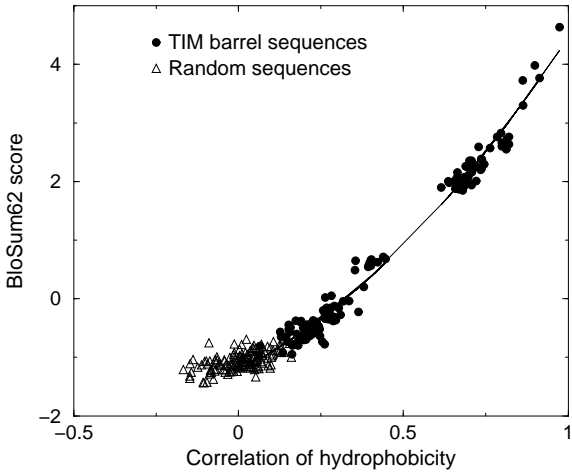}
}
\caption{For each pair of sequences, we compare the HC score and the BloSum62
score. Circles represent pairs of sequences with TIM barrel fold; triangles
represent random sequences with the same length and the same number of gaps as
the true sequences. Positions with gaps are always less than 20\% of the
sequence, and gaps are omitted in the calculation of the scores. The line shows
a quadratic fit.}
\label{fig:blosum}
\end{figure}

We consider $136$ pairs of sequences with the TIM barrel fold, as listed in the
FSSP database, that can be aligned for more than 80\% of their length, and that
have less than 90\% sequence identity. For each pair, we compare in
Fig.~\ref{fig:blosum} the similarity score measured through the BloSum62 score
and the HC score described above. As seen from the figure, the correlation
between the two scores is very strong, despite the fact that the HC score uses
only $20$ parameters, whereas the BloSum matrix consists of $210$ parameters.
None of the two scores is able to recognize all TIM barrel pairs with respect
to pairs of random sequences. These results confirm the reported correlation
between the HP of even distantly related proteins with the same fold
\cite{Sweet83}, therefore supporting the idea that the optimal HP represents
the fingerprint of the protein fold.

Also in this case, one may ask whether the correlation can be improved using
optimized parameters. To address this question, we maximized the mean
$r({\mathbf{h}},{\mathbf{h}}')$ over the set of pairs of sequences with TIM
barrel fold used above (see Methods). The optimization yielded a marginal
improvement, from 0.497 to 0.511, and the optimal parameters were almost
identical to the former ones (correlation coefficient 0.97, see
Table~\ref{tab:hydro}), whereas they departed further from the parameters
obtained by maximizing the mean $r({\mathbf{c}},{\mathbf{h}})$ over
single-domain PDB structures (correlation coefficient 0.93). Therefore, the
interactivity parameters given by the principal eigenvector of our interaction
matrix are close to optimal with respect to both sequence to structure and
sequence to sequence comparison.

\subsection*{Vectorial Protein Space}

The results presented here show that it is useful to represent both protein
sequences and structures as vectors in the same $N$-dimensional space. This
\textbf{Vectorial Protein Space} is endowed with a natural metric through the
correlation coefficient, Eq.~(\ref{eq:corr}). This representation provides a
unified framework for addressing three issues that are central in
bioinformatics: sequence to sequence alignments \cite{Smith}, structure to
structure alignments \cite{Holm96,Kohel01}, and sequence to structure
alignments for the purpose of protein structure prediction \cite{Jones92}.

Concerning \textbf{sequence to sequence} alignments, we saw that the HC score
$r({\mathbf{h}},{\mathbf{h}}')$, already proposed by Sweet and Eisenberg
\cite{Sweet83}, performs equally well as the BloSum62 score matrix
\cite{Henikoff92} despite having 20 parameters instead of 210. This result is
probably due to the fact that the HC score is context dependent: The HC score
for substituting residue $a$ in sequence ${\mathbf{A}}$ with residue $b$ in
sequence ${\mathbf{A}}'$ does not depend on $a$ and $b$ alone, but also on the
average and the variance of the HP in the two sequences.

\begin{figure}[ht]
\center{\leavevmode
\epsfxsize 7.6cm
\epsffile{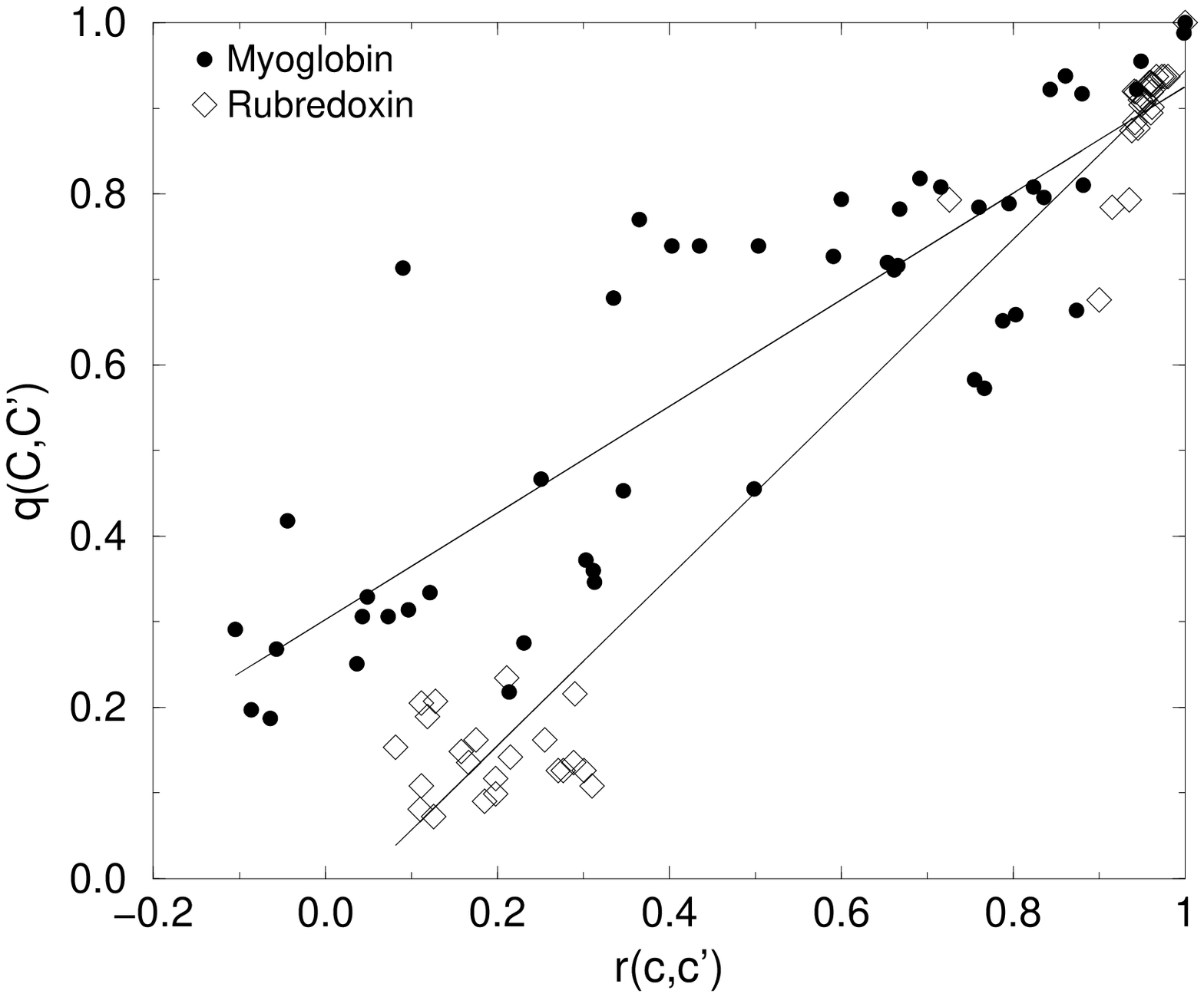}
}
\center{\leavevmode
\epsfxsize 7.6cm %11cm
\epsffile{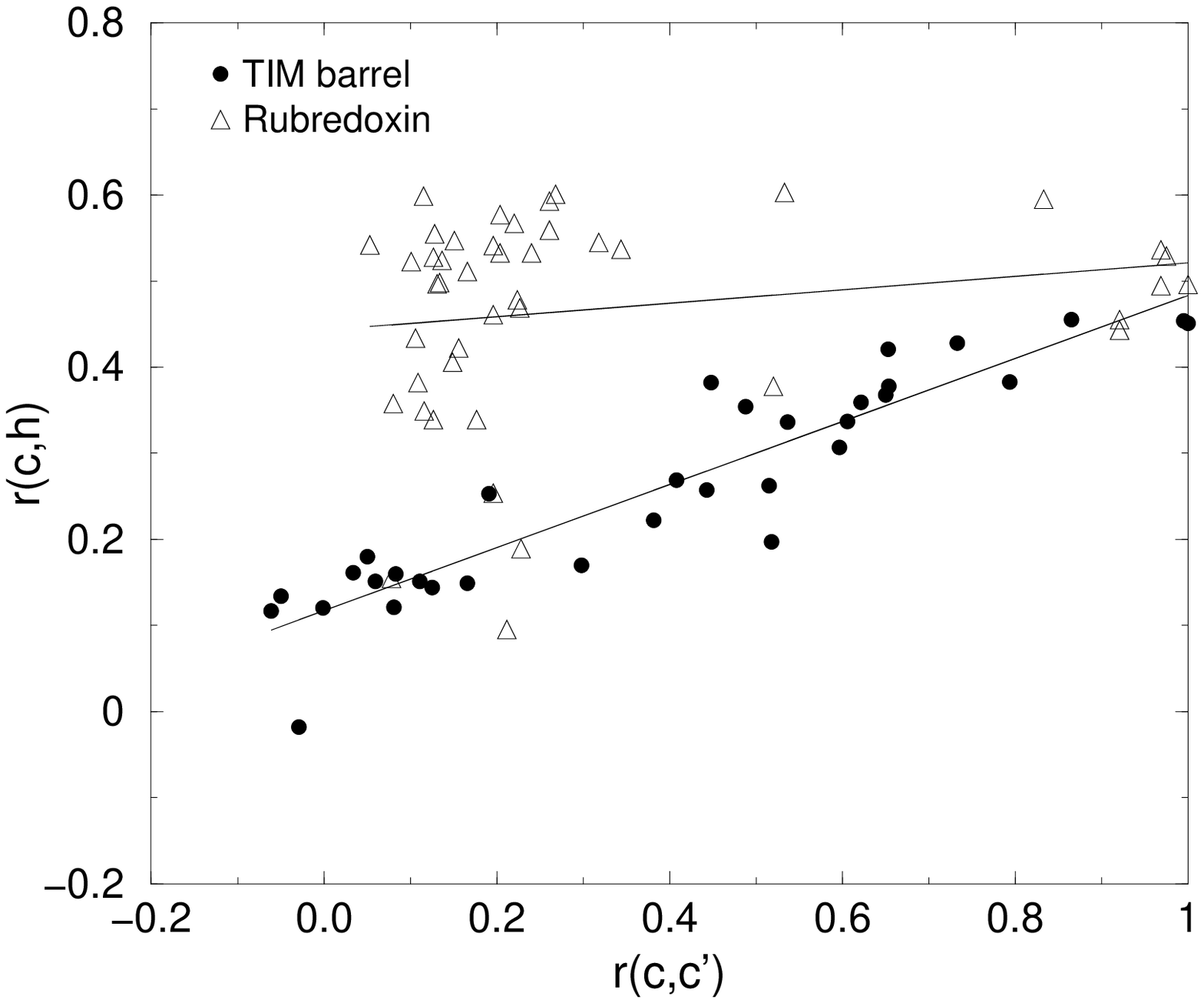}\hspace*{4mm}
}
\caption{Scatter plot of the contact overlap $q({\mathbf{C}},{\mathbf{C}}')$
(upper panel) and of the sequence to structure score
$r({\mathbf{c}},{\mathbf{h}})$ (lower panel) vs structural similarity measured
by $r({\mathbf{c}},{\mathbf{c}}')$. Here ${\mathbf{h}}$ indicates the HP of the
protein sequence, $\mathbf{c}'$ indicates the PE of the native state and
$\mathbf{c}$ indicates the PE of alternative states generated by threading with
gap penalties. In both cases, the two folds presenting largest and smallest
correlation are shown.}
\label{fig:qc}
\end{figure}

The $r({\mathbf{c}},{\mathbf{c}}')$ score for \textbf{structure to structure}
alignment is strongly correlated with the widely used contact overlap. When low
energy structures generated by threading are compared to the native one, the
two similarity measures have mean correlation coefficient $0.93$ for the seven
protein folds studied, i.e.\ they are almost equivalent. We show in
Fig.~\ref{fig:qc} the folds where the two distance measures have largest
(rubredoxin, 0.98) and smallest (myoglobin, 0.87) correlation coefficient.
Moreover, this score is strongly correlated with the effective energy function,
which is an important requisite of suitable measures of structural similarity
\cite{Wallin03}.

Finally, the HP-PE correlation score $r({\mathbf{c}},{\mathbf{h}})$ for
\textbf{sequence to structure} alignment is strongly correlated with the
effective free energy of the structure from which the PE is obtained and with
the similarity between this structure and the native one (see
Fig.~\ref{fig:qc}). Different from our effective free energy function
Eq.~(\ref{eq:energy}), this score does not assign the highest value to the
native structure with respect to decoys generated by threading. Nevertheless,
the highest scoring decoys are very similar to the native ones, particularly in
the case of large proteins. Moreover, all structures similar to the native tend
to have high $r({\mathbf{c}},{\mathbf{h}})$ score. Therefore, this score may be
useful for rapidly screening from a large database a restricted number of
candidate structures for more accurate fold recognition techniques, as expected
based on the success of sequence to structure comparison methods based on
profiles \cite{Bowie91,Jones92}.

%%%%%%%%%%%%%%%%%%%%%%%%%%%%%%%%%%%%%%%%%%%%%%%%%%%%%%%%%%%%%%%%%%%%%%%%%%%%

\section*{DISCUSSION}

The relationship between hydrophobicity and pair potentials has been
studied by various groups, including Li \textit{et al.} \cite{Li97},
Betancourt and Thirumalai \cite{Betancourt99} and more recently Cline
\textit{et al.} \cite{Cline02}. Here we show that the principal eigenvector
of the contact matrix of the native structure (PE), a global indicator of
protein structure, is positively correlated with the hydrophobicity profile
(HP) of its sequence and, more strongly, with the average HP of sequences
adopting the same fold.

The hydrophobicity scale used in this work was derived from the principal
eigenvector of the effective pair interaction matrix by Bastolla {\it et al.}
\cite{Bastolla01}. We have called it interactivity scale to underline the fact
that it embodies also other kinds of interactions besides the hydrophobic
effect. Its strong correlation with empirical hydrophobicity scales (the
correlation is $R=0.85$ with the octanole scale by Fauchere and Pliska) and the
demonstration that pair potentials are dominated by hydrophobicity justify our
simultaneous use of the name hydrophobicity profile. One should note, however,
that the word hydrophobicity is not used here with its strict biochemical
meaning.

After this paper was submitted, we became aware of a recent and interesting
paper where `buriability' parameters were derived for each amino acid from the
thermodynamic effects of site-directed mutagenesis \cite{Zhou04}.
Interestingly, these buriability parameters are almost identical to our
interactivity scales, with a correlation coefficient of 0.92 with the scale
derived from the principal eigenvector of the interaction matrix and a
correlation coefficient of 0.98 with the scale derived optimizing the HP-PE
correlation.

The HP can be useful for recognizing and aligning distantly related sequences,
as proposed by Sweet and Eisenberg \cite{Sweet83}, and for aligning sequences
and structures of related proteins. It is interesting, and perhaps surprising,
that the hydrophobicity parameters obtained from the principal eigenvector of
our interaction matrix \cite{Bastolla01} are almost optimal for both purposes,
since they almost maximize at the same time the correlation between distant
sequences sharing the same fold and the correlation between the PE of
single-domain globular proteins and the HP of their sequences. Therefore, the
results presented in this paper can help developing new bioinformatics methods
and algorithms to unify and perhaps improve different kinds of alignments.

 From Eq.~(\ref{eq:hydro}), one may expect a correlation between the PE and the
HP much stronger than the one observed, which is in the range 0.4 to 0.6. In
fact, the contact matrix with lowest effective free energy,
Eq.~(\ref{eq:energy}), is characterized by $r({\mathbf{c}},{\mathbf{h}})$ close
to one, since the contact matrix with $r({\mathbf{c}},{\mathbf{h}})=1$
minimizes the effective hydrophobic energy, Eq.~(\ref{eq:hydro}), for a fixed
value of the principal eigenvalue $\lambda_1$, expressing the number of
contacts per residue, and in turn Eq.~(\ref{eq:hydro}), gives the most
important contribution to the effective contact free energy. The ground state
of our protein model, which we identify with the native state, is the
protein-like structure having the lowest effective energy. By protein-like we
mean that local structure, dihedral angles, excluded volume interactions, local
electrostatic interactions, hydrogen bonds, etc., are distributed as in native
protein structures. These conditions are not enforced through the effective
energy function, but they are obtained constraining candidate structures to
fragments of protein crystal structures. By threading protein sequences against
the whole PDB database with a suitable gap penalty, we never found any PE of
protein contact matrices whose correlation with the HP was close to one. Such a
contact matrix would have lower effective free energy than the true native
state, whose correlation with the HP is of the order of $0.5$.

In the case of small proteins, as rubredoxin ($N=53$, see lower panel of
Fig.~\ref{fig:qc}), one can find alternative structures very different from the
native one with $r({\mathbf{c}},{\mathbf{h}})$ larger than for the native
structure, but still much smaller than one. But for proteins with more than 100
residues, such as for instance the TIM barrel ($N=247$, Fig.~\ref{fig:qc}) all
the structures with large correlation $r({\mathbf{c}},{\mathbf{h}})$ are very
similar to the native one. This results suggest that, for HP of natural
proteins, protein-like structures having $r({\mathbf{c}},{\mathbf{h}})$ close
to one do not exist, and open the question of why large regions of the
$N$-dimensional Vectorial Protein Space do not seem to contain vectors that are
the PE of the contact matrix of some protein-like structure.

The only moderate correlation between HP and PE has deep implications on
protein evolution. The requirement of a strong correlation would imply that
only sequences very similar to a protein fold (in the sense of the correlation
coefficient) are compatible with this fold. This would contrast with the
observation that the distribution of sequence identity for proteins adopting
the same fold approaches the distribution for random pairs of sequences
\cite{Rost97}. Further analysis will be needed to understand the relationship
between these two observations.

To gain further insight into the sequence-structure relationship, we have
defined the optimal HP of a given fold. Our simple model of protein folding,
Eq.~(\ref{eq:energy}), leads to the prediction that the optimal HP is strongly
correlated with the PE. Notice that this calculation also provides an
analytical solution to sequence design based on the effective energy function
Eq.~(\ref{eq:hydro}).

We expect that sequences sharing a given fold can not be too different from the
optimal HP, although this HP is never actually attained during protein
evolution. This interpretation is strongly supported by an analysis of
sequences derived from our SCN model of evolution. In this case, the PE of the
(fixed) reference structure is moderately correlated with the HP of individual
sequences, but it is very strongly correlated with the average of the HP over
all sequences. This suggests that our model of neutral evolution can be
described as a motion in sequence space around the optimal HP.

The same result is obtained considering sets of homologous sequences (PFAM) and
sets of sequences sharing the same fold (FSSP): The HP averaged over these
sequences is more strongly correlated with the PE than the HP of each
individual sequence. By means of this result, evolutionary information may give
a valuable contribution to the goal of predicting protein structure using the
PE as an intermediate step.

The fact that these correlations are much weaker for PFAM and FSSP families
than for the SCN model is not unexpected. First, functional constraints are
important in protein evolution, but they are not represented in the SCN model.
These may cause conservation of amino acids that are not optimal from a
thermodynamic point of view. Second, our structural model only considers
interactions between residues in the same chain, whereas, for several of the
proteins that we studied, interactions with cofactors and with other protein
chains play an important role. The presence of disulphide bridges and
iron-sulfur clusters is also problematic. Third, the FSSP and PFAM databases
contain much less sequences than those obtained through simulated evolution.
Decreasing the size of the SCN set, the correlation with the PE also decreases.

Part of the difference between observed and simulated protein evolution may
also be due to the fact that the free energy function in Eq.~(\ref{eq:energy})
is not sufficiently accurate to describe real proteins. Using this equation, we
are neglecting other kinds of interactions relevant for protein stability, such
as hydrogen-bonding, packing interactions, and entropic contributions to the
native free energy. It is possible that these neglected interactions are only
partially averaged out when summing over a large set of homologous proteins,
even if they contribute substantially to protein folding thermodynamics. This
may result in an increased, yet not perfect correlation between the average HP
and the PE. Alternatively, the hydrophobicity itself might be
context-dependent, for instance influenced by neighboring residues. This could
lead to what is indeed observed, {\it i.e.} a significant but not complete
correlation between HP and PE. However, we found that the parameters that
maximize the correlation between HP and PE are very strongly correlated with
the interactivity parameters obtained from the principal eigenvector of the
interaction matrix. This result supports the view that the interaction matrix
that we are using describes an important contribution to protein stability.

Another interesting property of the interactivity scales that we have derived
here, besides their expected strong correlation with protein folding
thermodynamics, is that the interactivity of orthologous protein also
correlates very strongly with properties of the genomes in which these proteins
are expressed (U. Bastolla {\it et al.}, submitted).

The relationship between HP and PE suggests that proteins of low sequence
similarity may share the same fold provided their HP are correlated with the
optimal HP. Therefore the optimal HP constitutes a common hydrophobic
fingerprint that characterizes a protein fold.

%%%%%%%%%%%%%%%%%%%%%%%%%%%%%%%%%%%%%%%%%%%%%%%%%%%%%%%%%%%%%%%%%%%%%%%%%%%%%

\section*{ACKNOWLEDGEMENTS}

UB is supported through the I3P Program of the CSIC (Spain), financed by the
European Social Fund. MV is supported by the Royal Society (UK). MP was
supported by the visitors program of the Max-Planck-Institut f\"ur Physik
komplexer Systeme, Dresden, Germany, during the first part of this work. We
thank an anonymous referee for useful comments.

%%%%%%%%%%%%%%%%%%%%%%%%%%%%%%%%%%%%%%%%%%%%%%%%%%%%%%%%%%%%%%%%%%%%%%%%%%%%%

%%%%%%%%%%%%%%%%%%%%%%% Tab. 2 %%%%%%%%%%%%%%%%%%%%%%%%%%%%%%%

\onecolumn
\textwidth 16.5cm
\oddsidemargin 0.true cm %
\evensidemargin 0.true cm %

%\begin{table*}[t]
%\rotate[l]{\rotate[r]{(Caption on next page)}
\begin{tabular}{|l|l|c|c|c|c|c|c|}\hline
&&&&&&&\\
Protein & PDB id. & $N$ &
$r({\mathbf{c}},{\mathbf{h}}_{\mathrm{PDB}})$ &
$\overline{r({\mathbf{c}},{\mathbf{h}}_{\mathrm{PFAM}})}$ &
$r({\mathbf{c}},\overline{{\mathbf{h}}_{\mathrm{PFAM}}})$ &
$r({\mathbf{c}},\overline{{\mathbf{h}}_{\mathrm{FSSP}}})$ &
$r({\mathbf{c}},\overline{{\mathbf{h}}_{\mathrm{SCN}}})$ \\
&&&&&&&\\
\hline
\hline
rubredoxin & \texttt{1iro}/\texttt{1brfA} &
%(meso-/thermophilic)
$53$  & $0.496$ & $0.465$ & $0.602$ & $0.599$ & $0.987$ \\
cytochrome~c                    & \texttt{451c}                  &
$82$  & $0.500$ & $0.491$ & $0.573$ & $0.684$ & $0.962$ \\
ribonuclease                    & \texttt{7rsa}                  &
$124$ & $0.431$ & $0.400$ & $0.491$ & $0.498$ & $0.965$ \\
lysozyme                        & \texttt{3lzt}                  &
$129$ & $0.531$ & $0.544$ & $0.649$ & $0.627$ & $0.949$ \\
myoglobin                       & \texttt{1a6g}                  &
$151$ & $0.399$ & $0.352$ & $0.472$ & $0.465$ & $0.966$ \\
ubiquitin conjugating enz.    & \texttt{1u9aA}               &
$160$ & $0.451$ & $0.450$ & $0.593$ & $0.567$ & $0.943$ \\
TIM barrel                      & \texttt{7timA}               &
$247$ & $0.466$ & $0.404$ & $0.486$ & $0.584$ & $0.970$ \\
\hline
\end{tabular}
%}
\begin{table*}[t]
\caption{Correlation coefficients between the PE and the HP,
$r({\mathbf{c}},{\mathbf{h}})$ for seven protein folds of different length $N$.
In column 4, the HP is obtained from the PDB sequence from which the PE is
calculated. In column~5, the $r$ is averaged over different PFAM sequences.
Single-sequence $r$'s are calculated using the alignments between the PDB
sequence and sequences from the same PFAM family. Mean values do not differ
significantly from those for the original PDB sequence. Similar values are
obtained averaging $r$ over sequences in the FSSP and SCN families (not shown).
In the remaining columns, the mean HP is obtained by averaging the HP of
sequences in the same family of the PFAM database (column~6), the FSSP database
(column~7), and the SCN database (column~8), respectively. Most values of the
PFAM and FSSP databases are very similar to each other and both higher than the
value of $r$ for individual sequences, but significantly smaller than the
corresponding values for the SCN database. Note that for the correlation
coefficient based on the average HP, sites containing cysteine residues are not
included, as they form pairwise disulphide bridges (which are very poorly
represented through the hydrophobic energy) and are strictly conserved in our
evolutionary model.}
\label{tab:summary}
\end{table*}

%%%%%%%%%%%%%%%%%%%%%%%%%%%%%%%%%%%%%%%%%%%%%%%%%%%%%%%%%%%%%%%%%%%%%%%%%%%%

\end{document}